\documentstyle[twoside,fleqn,espcrc2,psfig]{article}
\title{%
Simulation of $n_f =3$ QCD by Hybrid Monte Carlo
}
\author{Tetsuya Takaishi\address{Hiroshima University of Economics,
 Hiroshima, 731-0192, JAPAN}\thanks{Presented by T. Takaishi at LAT00.}
and Philippe de Forcrand\address{Inst. f\"ur Theoretische
    Physik, ETH H\"onggerberg, CH-8093 Z\"urich, Switzerland}
\address{CERN, Theory Division, CH-1211 Gen\`eve 23, Switzerland} }

\def\Journal#1#2#3#4{{#1} { #2} (#4) #3}

\def\NPB{{\em Nucl. Phys.} B}
\def\PLB{{\em Phys. Lett.}  B}

\def\PRD{{\em Phys. Rev.} D}

\def\ra{\rightarrow}

\def\be{\begin{equation}}
\def\ee{\end{equation}}
\def\bea{\begin{eqnarray}}
\def\eea{\end{eqnarray}}

\begin{document}
\begin{abstract}
Simulations of odd flavors QCD
can be performed in the framework of the hybrid Monte Carlo algorithm
where
the inverse of the fermion matrix is approximated by a polynomial.
In this exploratory study we perform three flavors QCD simulations.
We make a comparison of the hybrid Monte Carlo algorithm and the R-algorithm
which also simulates odd flavors systems but has step-size errors.
We find that results from our hybrid Monte Carlo algorithm are in agreement
with those from the R-algorithm
obtained at very small step-size.
\end{abstract}

\maketitle

\section{Introduction}

Recent lattice QCD simulations include effects of dynamical fermions.
Due to the algorithmic limitation of the standard
Hybrid Monte Carlo (HMC) algorithm \cite{HMC}, those simulations are
limited to even numbers of degenerate flavors.
In order to include dynamical effects correctly, simulations of
QCD with three flavors (u,d,s quarks) are desirable.
Simulations with an odd number of flavors can be performed using
the R-algorithm \cite{Ralg}. This algorithm, however, is not exact:
it causes systematic errors of order $\Delta\tau^2$, where $\Delta\tau$
is the step-size of the Molecular Dynamics evolution. A careful
extrapolation to zero step-size is therefore needed to obtain exact
results. Nevertheless, it is common practice to forego this extrapolation
and to perform simulations with a single step-size chosen small enough
that the expected systematic errors are smaller than the statistical ones.
We want to point out that there is an alternative to the R-algorithm,
which gives arbitrarily accurate results without any
extrapolation\cite{other}.

L\"uscher proposed a local algorithm, the so-called "Multiboson algorithm"
\cite{Luscher},
in which the inverse of the fermion matrix
is approximated by a suitable Chebyshev polynomial.
Originally he proposed it for two flavors QCD. Bori\c ci and
de Forcrand \cite{BPh} noticed that the determinant of a fermion matrix can
be
written in a manifestly positive way using a polynomial approximation, so
that
one can simulate odd flavors QCD with the multiboson method.
Indeed, using this method, one flavor QCD was simulated successfully
\cite{NF1}.
The same polynomial approximation can be applied for the HMC \cite{FFMC}.
Actually, in the development stage of Ref.\cite{NF1},
one flavor QCD was also simulated by HMC and
it was confirmed that the two algorithmically different methods ---
multiboson and HMC --- give the same
plaquette value \cite{nf1HMC}.
Here we give the formulation of the HMC
algorithm with odd flavors and perform $n_f=3$ QCD simulations.
Then we compare our results with those of the R-algorithm.

\section{Formulation}

\subsection{$n_f=2$}
The application of L\"uscher's idea \cite{Luscher} to $n_f =2$ QCD HMC was
first made by
the authors of Ref.\cite{FFMC},  and later by \cite{Jansen}.
The lattice QCD partition function with $n_f=2$ degenerate quark flavors is
given by
\be
Z=\int dU \det D^{2} \exp(-S_{gauge}),
\ee
where $D$ is the fermion matrix and in this study we use Wilson fermions.
In the formulation of the HMC algorithm, $\det D^{2}$ is treated as
\be
\det D^2 \sim \int d{\phi}^\dagger d{\phi}
\exp(- {\phi}^\dagger D^{\dagger -1}D^{-1} \phi),
\label{eq:det2}
\ee
where the $\gamma_5$ hermiticity of the fermion matrix $D$, i.e. $D=\gamma_5
D^\dagger \gamma_5$, is used.

Introducing momenta $P$ conjugate to the link variables $U$, the partition
function is rewritten as
\be
Z=\int dU dP \exp(-H),
\ee
where the Hamiltonian $H$ is defined by
\be
H=\frac{1}{2}P^2 +S_{gauge}   +  {\phi}^\dagger {D}^{\dagger -1}{D}^{-1}
{\phi} .
\label{eq:standard-H}
\ee
This Hamiltonian is used for the Molecular Dynamics (MD) simulation of the
standard HMC algorithm.
Eq.(\ref{eq:standard-H}) has a computational difficulty in MD simulations
since one must solve $x=D^{-1}\phi$ type equations which in general take a
large amount of computational time for
a large fermion matrix and/or for a small quark mass.

Following L\"uscher \cite{Luscher},
the inverse of $D$ can be approximated by a polynomial:
\be
1/D\approx P_n(D) \equiv \prod_{k=1}^{n}(D-Z_k),
\ee
where $Z_k$ are the roots of the polynomial $P_n(D)$.
We choose $Z_k=1-\exp(i~2\pi k/(n+1))$.

Replacing $D^{-1}$ in eq.(\ref{eq:standard-H}) by $P_n(D)$ we obtain an
approximate Hamiltonian,
\be
H_n=\frac{1}{2}P^2 +S_{gauge}   +  {\phi}^\dagger P_n(D)^\dagger P_n(D)
{\phi} .
\ee
An advantage of using $H_n$ is that no solver calculation is  required in
the MD evolution.
Instead, one needs $n$ multiplications by the matrix $D$.

$H_n$ does introduce some systematic errors from the polynomial
approximation.
For the $n_f$=even case, however, these errors are easily corrected
by using the exact Hamiltonian of eq.(\ref{eq:standard-H}) at the Metropolis
step\cite{FFMC}.
This guarantees that configurations will be distributed according to
the exact measure $\propto \det D^2$, for {\em any} polynomial $P_n(D)$.

However, the domain of convergence of $P_n(D)$ is bounded by a circle
centered at
$(1,0)$ which goes through the origin.
If all eigenvalues of $D$ fall inside this domain, $P_n(D)$ converges
exponentially.
Otherwise, $P_n(D)$ does not converge, which may happen for some exceptional
configurations.
Our algorithm will tend to reject these configurations at the Metropolis
step,
leading to extremely long autocorrelation times.
This domain of convergence can be changed by adopting another approximating
polynomial. However, the origin must be excluded. Together with
connectedness
and conjugate symmetry of the spectrum, this implies that the real negative
axis is always excluded from the domain of convergence for any polynomial.
Configurations with real negative Dirac eigenvalues will be rejected by our
polynomial algorithm.

\subsection{$n_f=1$}

After invention of the multiboson algorithm,
Bori\c ci and de Forcrand \cite{BPh} noticed that a single $\det D$ can be
treated in a manifestly positive way
and an $n_f=1$ multiboson simulation was performed to study thermodynamics
of
$n_f=1$ QCD \cite{NF1}.

As before, the inverse of the fermion matrix $D$, using a polynomial of
degree $2n$, is approximated as \cite{Luscher,BPh}
\be
1/D\approx\prod_{k=1}^{2n}(D-Z_k),
\label{eq:nf1det}
\ee
where $Z_k=1-\exp(i~2\pi k/(2n+1))$.
Noticing that the $Z_k$'s come in complex conjugate pairs,
eq.(\ref{eq:nf1det}) is rewritten as
\be
1/D\approx\prod_{k=1}^{n}(D-\bar{Z}_k)(D-Z_k).
\ee
Using the $\gamma_5$ hermiticity of the fermion matrix,
we find that $\det (D-\bar{Z}_k)=\det(D-Z_k)^\dagger$.
Thus the determinant of $D$ is written as
\be
\det (D) \sim \det ( T_n^\dagger(D)T_n(D))^{-1},
\ee
where $T_n(D)\equiv\prod_{k=1}^n(D-Z_k)$,
and then we obtain
\begin{equation}
\det (D) \sim \int d\phi^\dagger d\phi \exp(-\phi^\dagger
T_n^\dagger(D)T_n(D)\phi).
\label{eq:det1}
\end{equation}
The term $\phi^\dagger  T_n^\dagger(D)T_n(D) \phi$
is manifestly positive.
Then we may define the Hamiltonian of $n_f=1$ QCD as
\be
H=\frac{1}{2}P^2 +S_{gauge}    +  \phi^\dagger  T_n^\dagger(D)T_n(D)\phi.
\label{eq:nf1-H}
\ee
With this Hamiltonian there is no difficulty to perform HMC algorithm.
To improve efficiency and accuracy, one may use a polynomial of lower degree
$n$ during
the Molecular Dynamics trajectory, and a much higher degree $m \gg n$ for
the Metropolis step \cite{NF1,Montvay}.
The domain of convergence of the approximation eq.(\ref{eq:det1}) is the
same
as for $n_f=2$. Exceptional configurations for which eigenvalues fall
outside this domain will likewise be rejected at the Metropolis step.
A further difficulty is that the sampled measure is
$\propto \det ( T_m^\dagger(D)T_m(D))^{-1}$, which for exceptional
configurations differs from the desired $\det D$, increasingly so with $m$.

\begin{figure}[h]
\vspace{-5mm}
\psfig{figure= 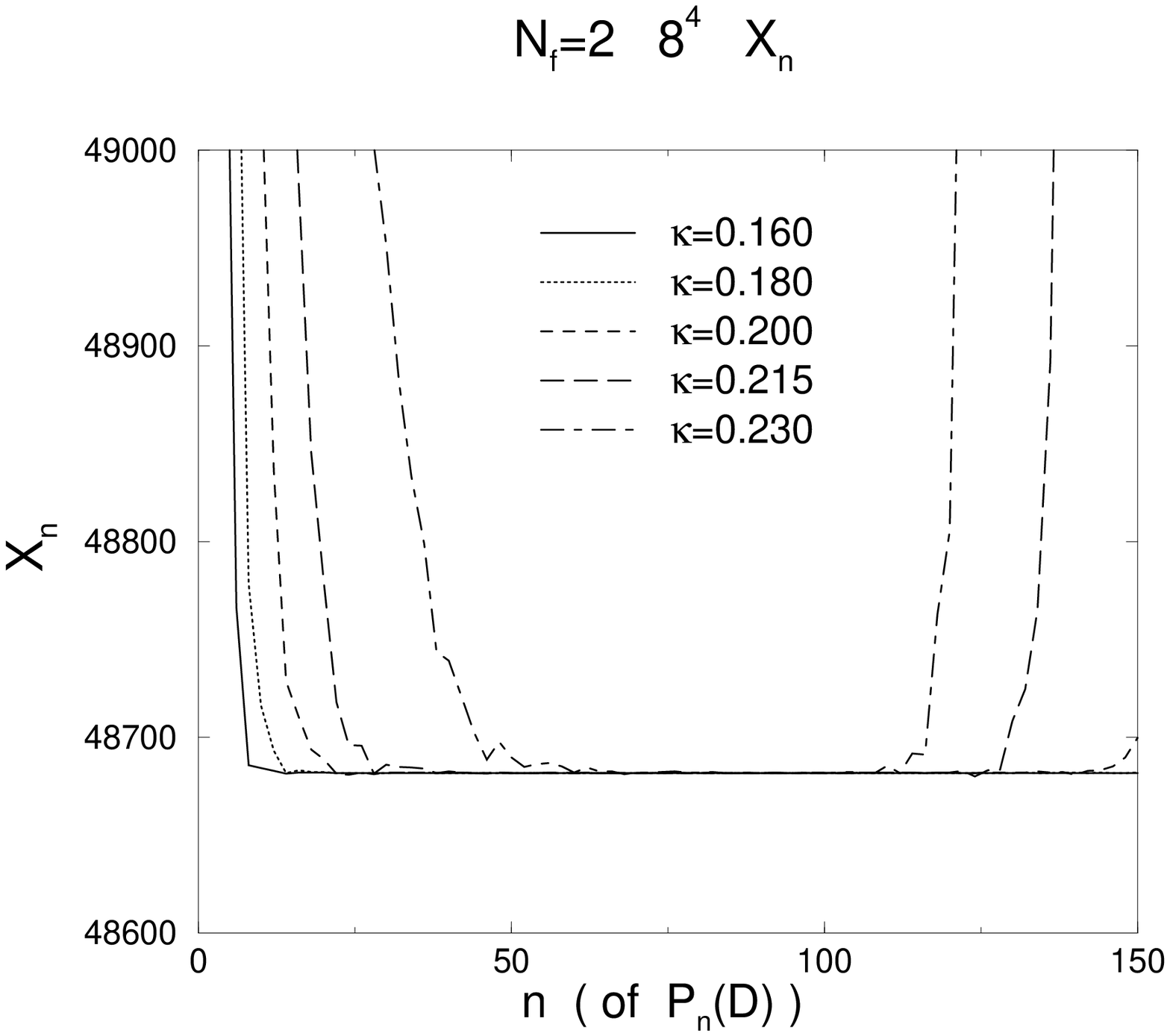,height=5.5cm}
\vspace{3mm}
\psfig{figure= 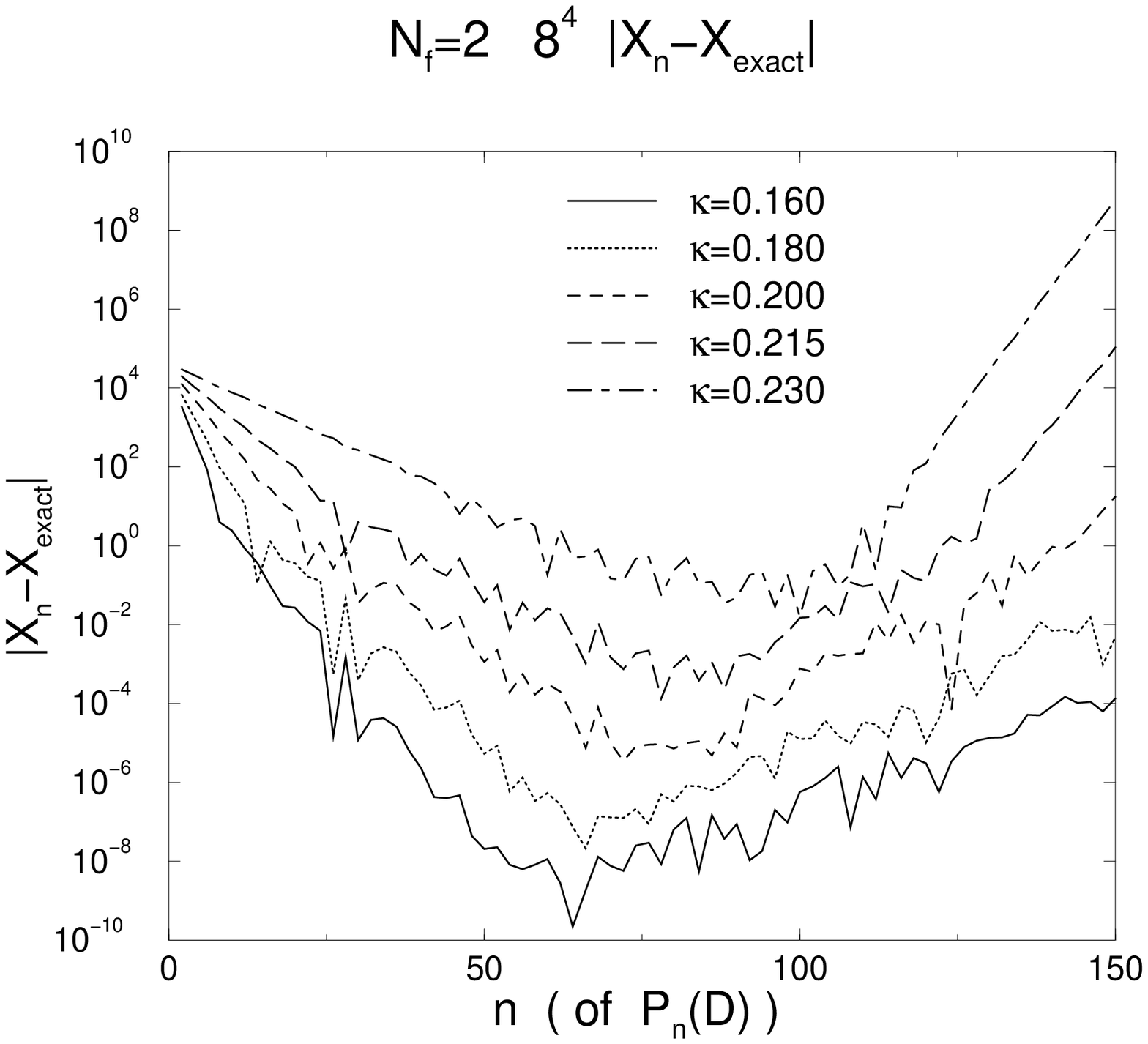,height=5.5cm}
\vspace{-10mm}
\caption{
(top): $X_n$ versus degree $n$; (bottom):$|X_n -X_{exact}|$ versus degree
$n$.
}
\vspace{-8mm}
\end{figure}

\subsection{$n_f=2+1$}

The partition function of $n_f=2+1$ QCD is given by
\be
Z=\int dU \det\tilde{D}^{2}\det D \exp(-S_{gauge}),
\ee
where the notations $\tilde{D}$ and $D$ are introduced to distinguish the
two different quark masses.
Using eq.(\ref{eq:det2}) for $\det\tilde{D}^{2}$ and eq.(\ref{eq:det1}) for
$\det D$,
\be
\begin{array}{l}
{\det\tilde{D}^{2}\det D} \\
{ \sim  \int  d\tilde{\phi}^\dagger d\tilde{\phi}
d\phi^\dagger d\phi
\exp(- \tilde{\phi}^\dagger {\tilde{D}^{\dagger -1}}{\tilde{
D}}^{-1} \tilde{\phi}-H_{T})},
\end{array}
\ee
where $H_{T}\equiv \phi^\dagger  T_n^\dagger(D)T_n(D)\phi$.
We define $n_f=2+1$ Hamiltonian by
\be
H=\frac{1}{2}P^2 +S_g   +  \tilde{\phi}^\dagger
{\tilde{D}^{\dagger -1}}{\tilde{D}}^{-1} \tilde{\phi}
+H_{T}.
\label{eq:H}
\ee

Two remarks are in order: $(i)$ as for $n_f=1$, one could use during the MD
trajectory a polynomial of lower degree than for the Metropolis step;
$(ii)$ the two bosonic fields $\phi$ and $\tilde{\phi}$ could be
replaced by a single one, with action
$\phi^\dagger  T_n^\dagger(D) {\tilde{D}^{\dagger -1}}{\tilde{D}}^{-1}
T_n(D) \phi$.
For simplicity, in this exploratory study we use two distinct bosonic fields
and a single approximating polynomial.

\section{Convergence}

\begin{figure}[t]
\psfig{figure= 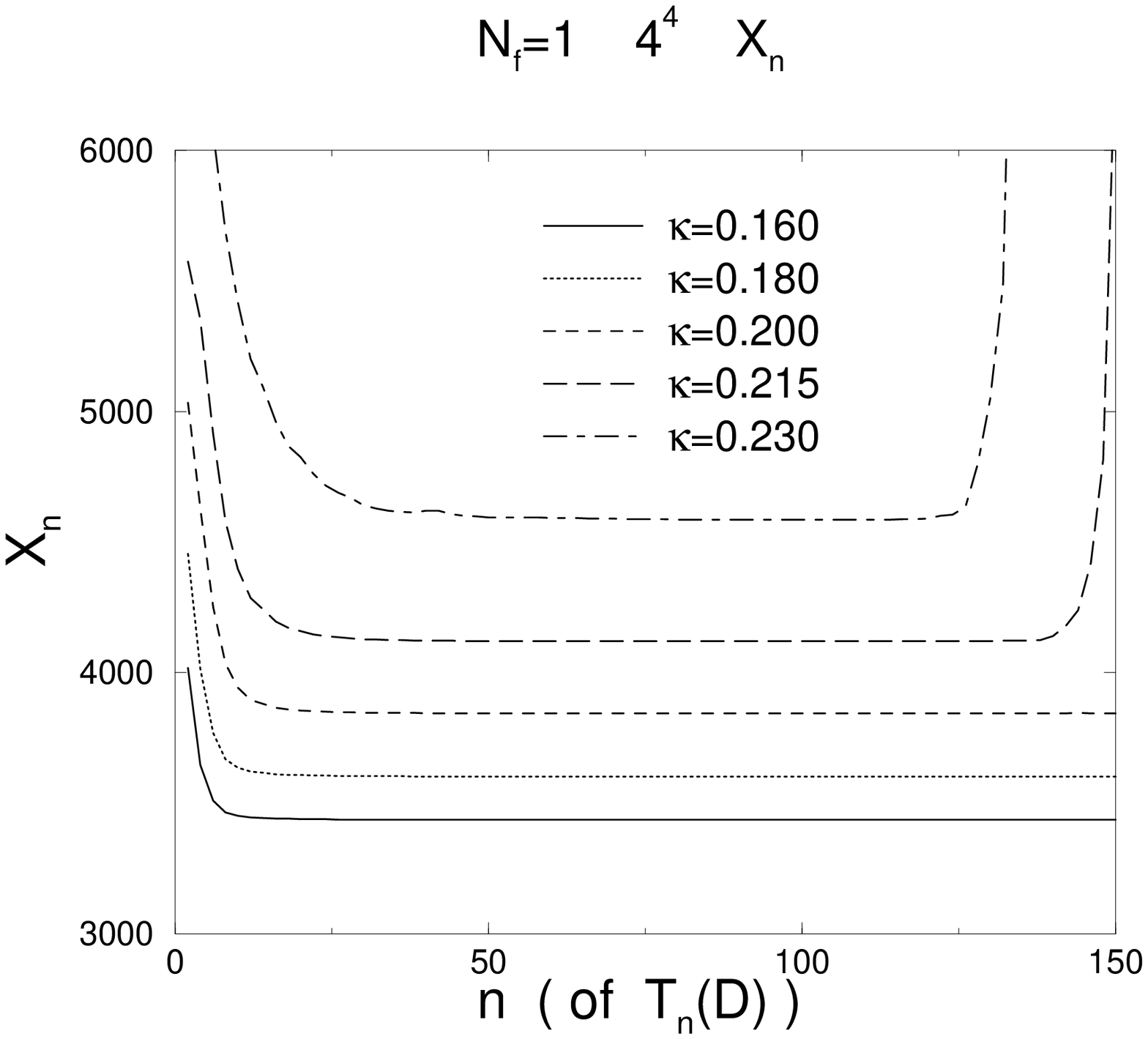,height=5.5cm}
\vspace{3mm}
\psfig{figure= 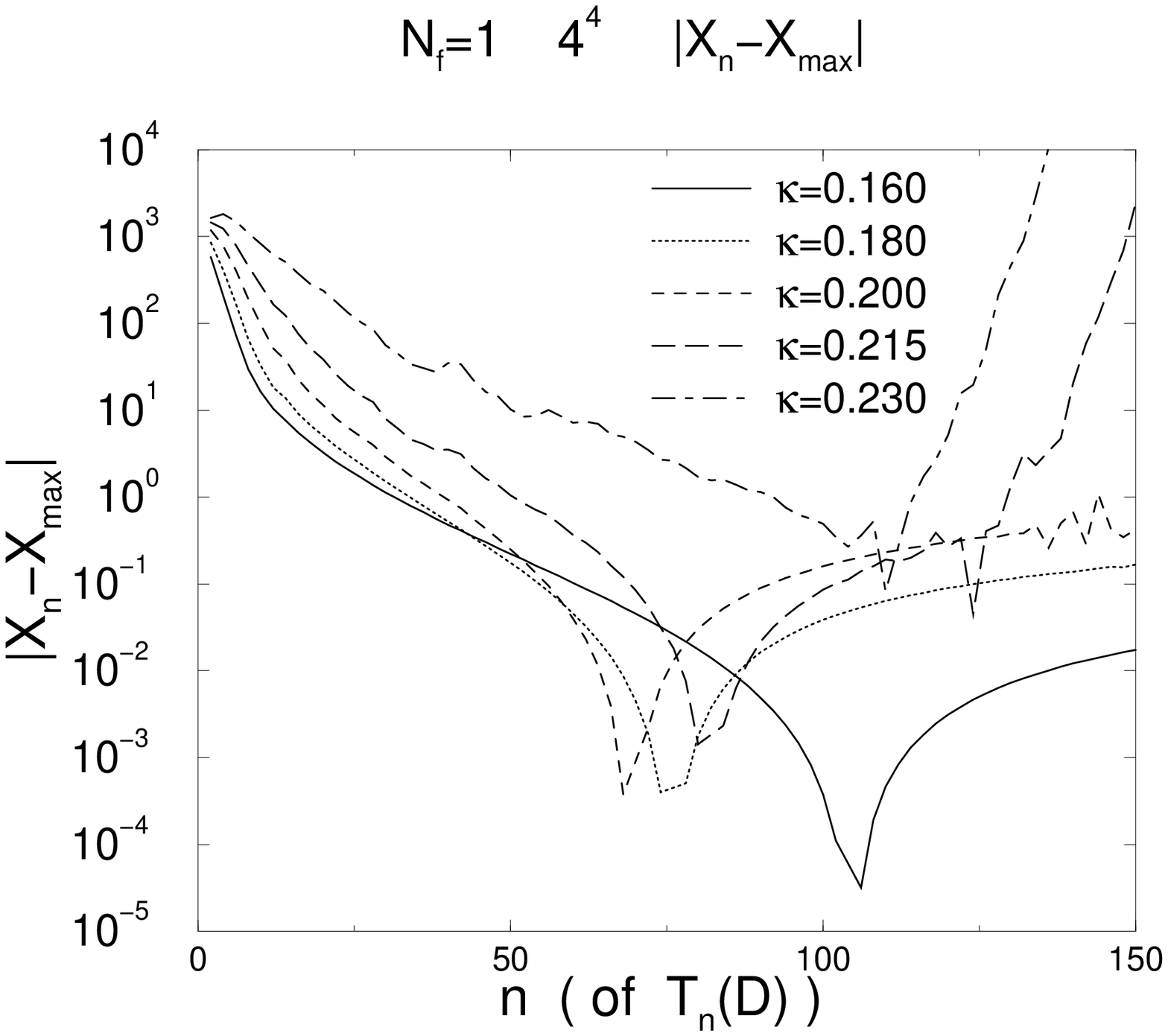,height=5.5cm}
\vspace{-10mm}
\caption{
(top): $X_n$ versus degree $n$; (bottom):$|X_n -X_{max}|$ versus degree $n$.
}
\end{figure}

\subsection{$n_f=2$}

To see the rate of convergence of $P_n(D)$,
we calculate the quantity $X_n=\phi^\dagger  P_n^\dagger(D)P_n(D)\phi$.
In the limit $n \ra \infty$, $X_n$ goes to $X_{exact}\equiv \phi^\dagger
{D^\dagger}^{-1}D^{-1}\phi$.
First, we choose $X_{exact}=\eta^\dagger  \eta$ where $\eta$ is a random
gaussian vector.
Then the vector $\phi$ is set to $\phi\equiv D\eta$.
The accuracy of $X_n$ is measured by the difference between $X_n$ and
$X_{exact}$.
We use a random gauge configuration.
Figure 1:(top) shows $X_n$ versus the degree $n$ for different quark
masses.
Here the same $\eta$ is used for each calculation of $X_n$.
$X_n$ converges to one value as $n$ increases,
but at high degree $n$, $X_n$ diverges,
which can be understood due to the rounding errors
of our computer, where calculations are performed with 64-bit accuracy.
Figure 1:(bottom) shows the accuracy of $X_n$ by $|X_n -X_{exact}|$.
Exponential convergence is seen for each quark mass,
but the rate of convergence is slow for small quark masses.

\subsection{$n_f=1$}

We do the same analysis as for $n_f =2$,
but for $n_f=1$, the value of $X_{exact}$ is not known.
So we calculate the quantity $X_n=\phi^\dagger  T_n^\dagger(D)T_n(D)\phi$,
where the vector $\phi$ is a gaussian random vector, and
we use a random gauge configuration.
We assume that
$X_n$ goes to a certain value in the limit of $n \ra \infty$.
Figure 2:(top) shows $X_n$ as a function of degree $n$.
$X_n$ seems to converges to a certain value when the degree $n$ increases.
At high degree $n$, $X_n$ diverges as in the case of $n_f=2$.

To see the rate of convergence, we calculate $|X_n-X_{max}|$ where
$X_{max}$ is defined by $X_{max}=X_m$, $m \gg n$.
Due to the rounding errors, we can not take $m$ very large.
We take a maximum number $m$ where the rounding errors still do not appear.
Figure 2:(bottom) shows $|X_n-X_{max}|$ as a function of degree $n$.
The dips seen in the figure are just due to the fact that
at those points $X_n=X_{max}\equiv X_m$.
The convergence seems to be exponential, but the rate of convergence is
slow for small quark masses as in the $n_f=2$ case.

\begin{figure}[t]
\psfig{figure=  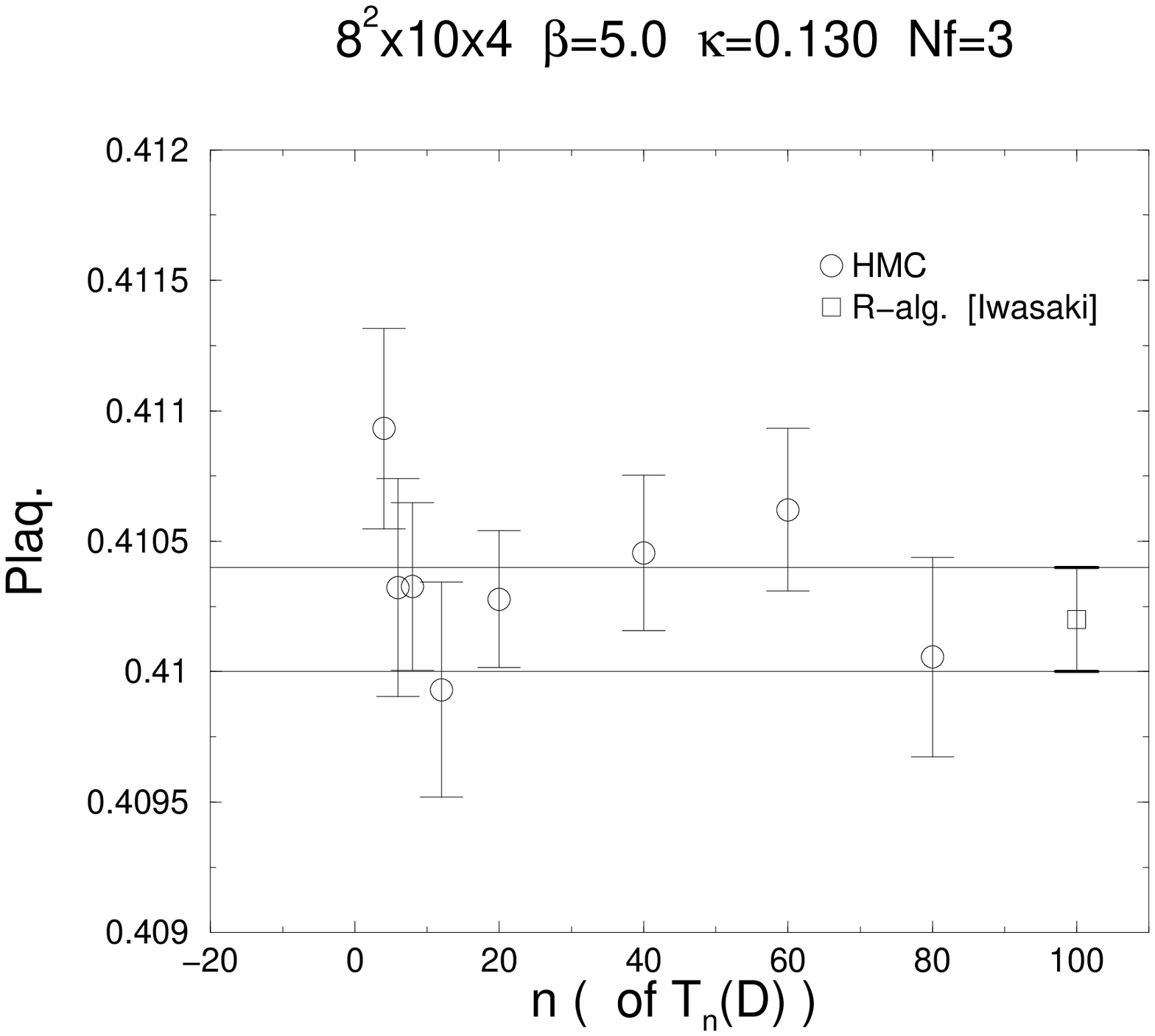 ,height=5.5cm}
\vspace{3mm}
\psfig{figure=  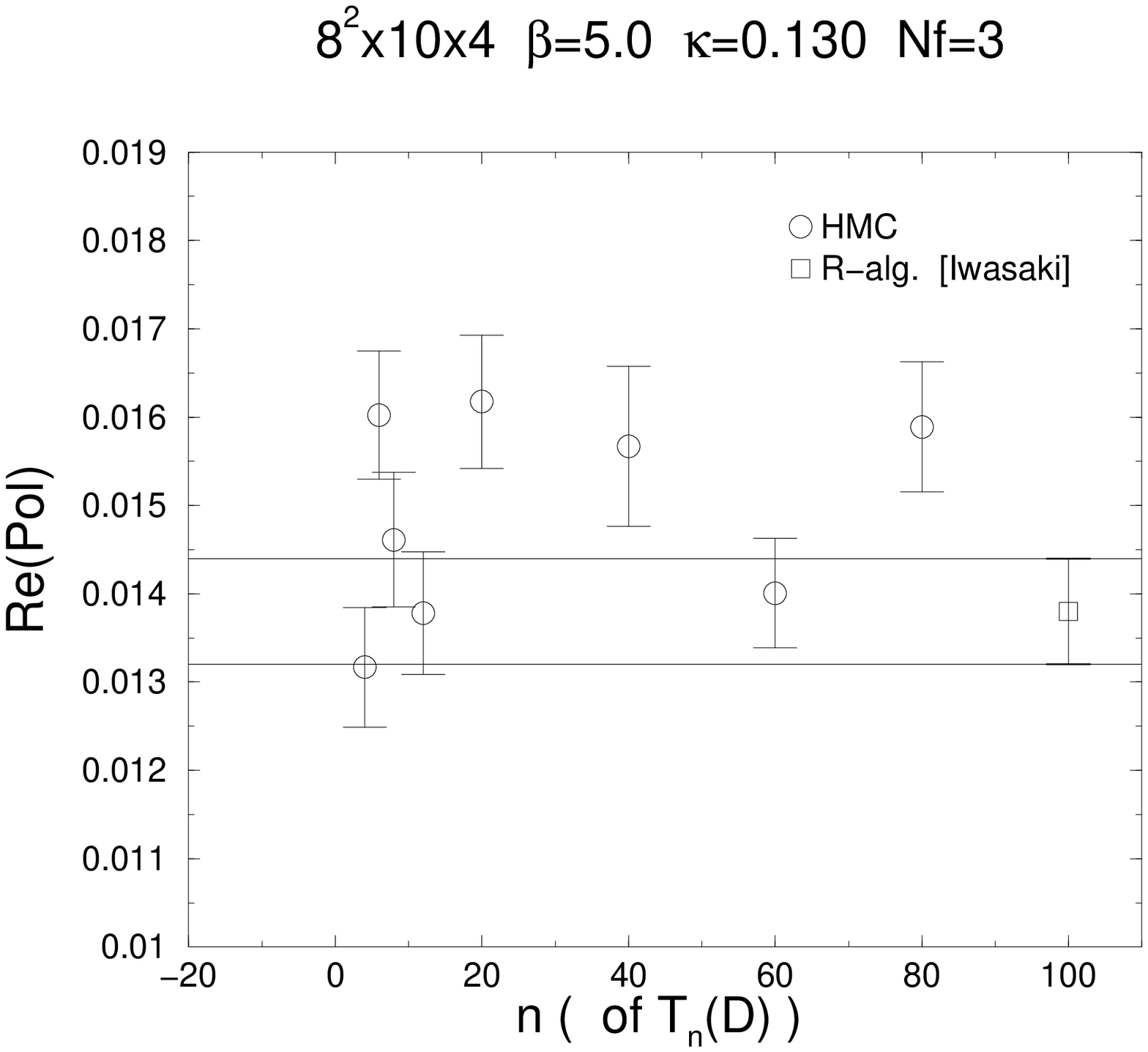 ,height=5.5cm}
\vspace{-10mm}
\caption{
(top): Plaquette of $n_f=3$ flavor QCD on an $8^2\times 10\times 4$
lattice at $\beta=5.0$ and at $\kappa =0.130$ as a function of degree $n$;
(bottom): Real part of Polyakov loop.
}
\end{figure}

\section{Simulations}
Simulations of three flavors QCD are performed on an $8^2\times 10 \times 4$
lattice at $\beta=5.0$
with $\kappa=0.130$ and $0.160$.
We measure the plaquette and Polyakov loop varying
the degree $n$ and compare them with those from the R-algorithm
obtained with a step-size $\Delta\tau=0.01$ \cite{Iwasaki}.
Figures 3 and 4:(top) show the plaquette as a function of $n$ at
$\kappa=0.130$ and 0.160, respectively.
Except for very small $n$,
the results from the HMC algorithm agree with those from the R-algorithm
within statistical errors.
Results of the Polyakov loop are shown in Figures 3 and 4:(bottom).
Except for a small discrepancy seen in Figure 3, the results from the HMC
algorithm
are in agreement with those from the R-algorithm.
Note that convergence is not monotonic in $n$.

\section{Conclusions}
We formulated an odd-flavor HMC algorithm using a polynomial approximation.
Simulations of three flavors QCD were performed.  We found that the
plaquette values are consistent with those
from the R-algorithm at very small step-size.
In principle the HMC algorithm is able to simulate any flavors of QCD,
with arbitrary accuracy and without extrapolation
[as long as all Dirac eigenvalues are not real negative].
However the rounding errors should be under control
when we use a large lattice or/and small quark masses
where one may need a polynomial of high degree $n$ to achieve sufficient
approximation.

\begin{figure}[t]
\psfig{figure= 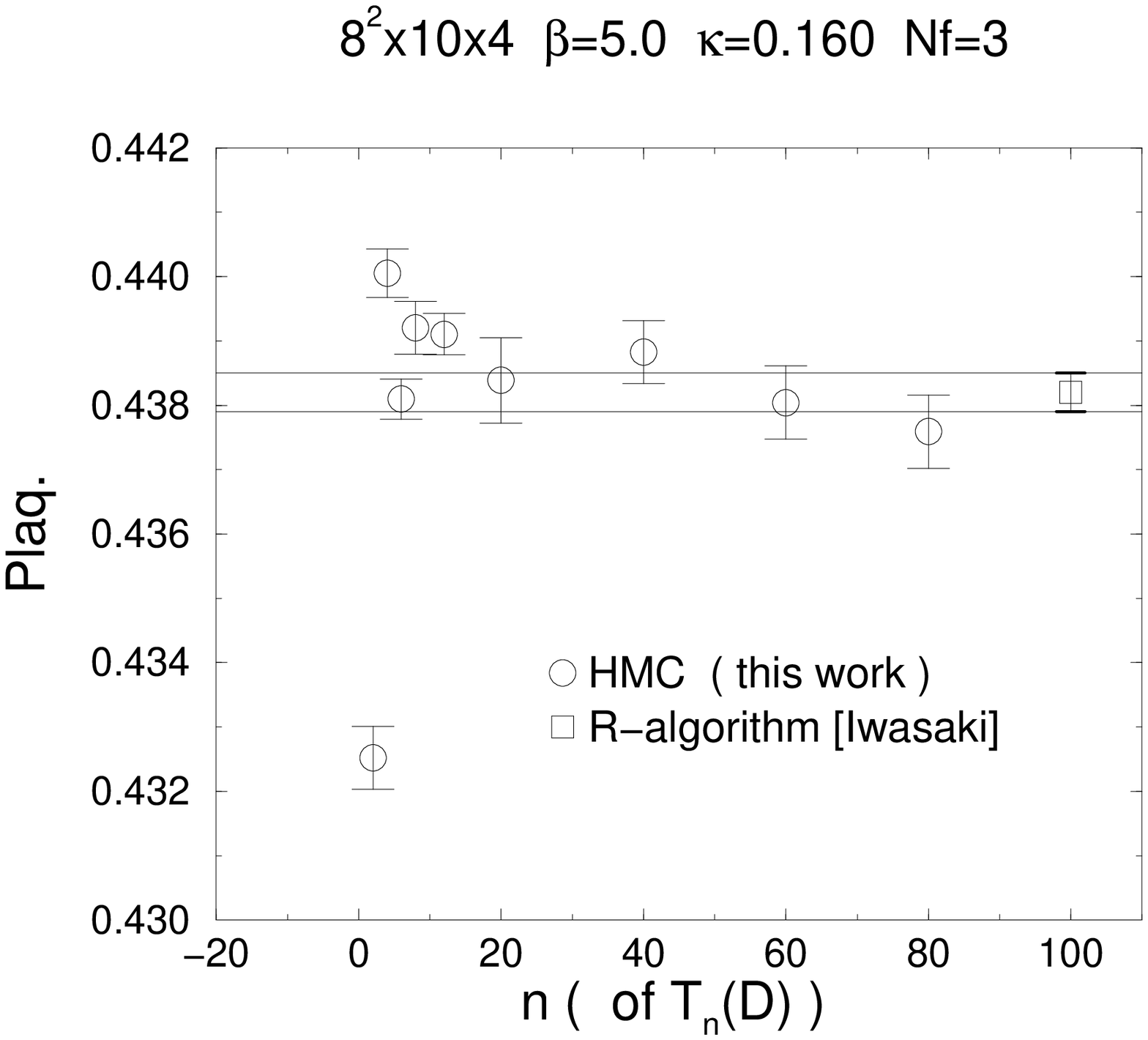,height=5.5cm}
\vspace{3mm}
\psfig{figure= 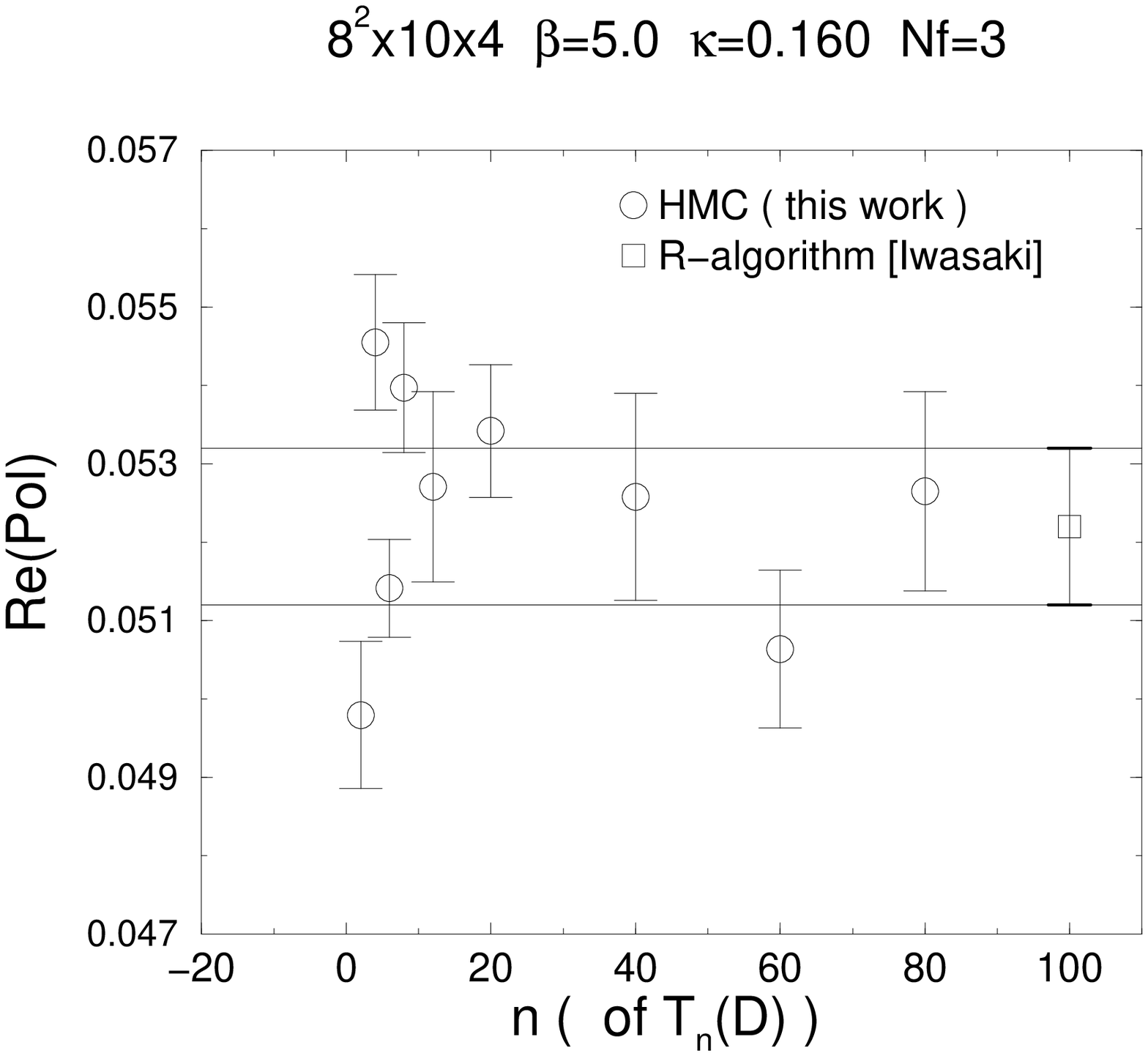,height=5.5cm}
\vspace{-10mm}
\caption{
(top): Plaquette of $n_f=3$ flavor QCD on an $8^2\times 10\times 4$
lattice at $\beta=5.0$ and at $\kappa =0.160$ as a function of degree $n$;
(bottom): Real part of Polyakov loop.
}
\end{figure}

\vspace{3mm}
This work was partially supported by the Minsitry of Education, Science,
Sports and Culture,
Grant-in-Aid, No.11740159.


\end{document}